\documentclass[twocolumn,nofootinbib]{revtex4}

\usepackage[usenames,dvipsnames]{color}
\usepackage{amsfonts,amssymb,amsmath,amsthm}
\usepackage{bm}
\usepackage[a4paper]{geometry}
\usepackage{pdflscape}
\usepackage{graphicx}
\usepackage{mathrsfs}

\newcommand{\ket}[1]{| #1 \rangle}

\newcommand{\bra}[1]{\langle #1 |}

\newcommand{\brast}[1]{\langle #1 |}
\newcommand{\braket}[2]{\langle#1|#2\rangle}

\newcommand{\avg}[1]{\langle#1\rangle}

\definecolor{darkred}{rgb}{.8,0,0}

\definecolor{darkblue}{rgb}{0,0,.7}

\newcommand{\rev}[1]{\widetilde{#1}}
\newcommand{\blade}[2]{{#1}_1 \wedge \ldots \wedge {#1}_{#2}}

\begin{document}

\title{Real spinors and real Dirac equation}

\author{V\'{a}clav Zatloukal}

\email[E-mail: ]{zatlovac@gmail.com}

\homepage[URL: ]{zatlovac.eu}

\affiliation{\vspace{3mm}
Faculty of Nuclear Sciences and Physical Engineering, Czech Technical University in Prague, \\
B\v{r}ehov\'{a} 7, 115 19 Praha 1, Czech Republic \\
}

\begin{abstract}
We reexamine the minimal coupling procedure in the Hestenes' geometric algebra formulation of the Dirac equation, where spinors are identified with the even elements of the real Clifford algebra of spacetime. This point of view, as we argue, leads naturally to a non-Abelian generalisation of the electromagnetic gauge potential.
\end{abstract}

\maketitle


\section{Introduction}

The real Clifford (or, geometric) algebra is a convenient tool to handle geometric objects, and study relations among them \cite{Clifford}. Greatly developed in \cite{HestenesSTA,Hestenes,DoranLas}, it has been proposed as a universal language for theoretical physics with a potential to bridge gaps between the classical and quantum domains.

In this article we draw attention to spinors, both in space and in spacetime, and advocate the approach of \cite{Hestenes1967}, in which they are regarded as elements of the even subalgebra of the \emph{real} Clifford algebra (see also \cite[Ch.\,8]{DoranLas} or \cite{Francis2005} for more recent treatment). This approach is reviewed in Sec.\,\ref{sec:RealCliff}, after a brief introduction to (real) Clifford algebras, and, in Sec.~\ref{sec:Relating}, related to a more traditional definition of spinors as members of complex vector spaces, hence providing a geometric interpretation for the Dirac matrices, and the imaginary unit $i$.

After these mathematical preliminaries, we focus, in Sec.\,\ref{sec:DiracEq}, on the Dirac equation of relativistic quantum mechanics. In the real spinor formalism, the corresponding reformulation, the real Dirac(-Hestenes) equation \cite{Hestenes1967,Hestenes1975}, is a relation within the real Clifford algebra of spacetime. 
We analyse this equation on a generic curved background, where tetrad (vierbein) fields are introduced to maintain invariance under a change of coordinates. 

The tangent spaces at every point of the spacetime manifold are isometrically mapped onto a `model' flat Minkowski spacetime (over which we construct a real Clifford algebra), and the local Lorentz transformations are understood as the ways to modify these mappings. In effect, the local Lorentz transformations act on both sides of all Clifford algebra elements, including the real spinors, so to maintain invariance of the real Dirac equation we introduce two collections of gauge fields: a spin connection for the action on the left, and another non-Abelian connection for the action on the right. The latter encompasses, and generalizes, the (Abelian) electromagnetic gauge potential. 

Proposing this non-Abelian extension of the electromagnetic field is the main purpose of the present article.

\section{Real Clifford algebras and real spinors}
\label{sec:RealCliff}

Let $V$ be an $n$-dimensional \emph{real} vector space, and $Q:V \rightarrow \mathbb{R}$ a non-degenerate quadratic form with signature $(p,q)$. The real Clifford algebra $\mathcal{C\ell}(V^{p,q})$, referred to as the geometric algebra in \cite{Clifford,HestenesSTA,Hestenes,DoranLas}, is the `freest' tensor algebra generated by $V$, subject to the condition \cite{Chevalley}
\begin{equation} \label{Square}
a^2 = Q(a) .
\end{equation}

Multiplication in this algebra, the Clifford (or geometric) product, is associative and distributive, but, in general, not commutative. For a pair of vectors $a$ and $b$, we can write
\begin{equation}
a b = a \cdot b + a \wedge b ,
\end{equation}
where $a \cdot b \equiv \frac{1}{2}(a b + b a)$ denotes the symmetric part of the Clifford product, while $a \wedge b \equiv \frac{1}{2}(a b - b a)$ denotes its antisymmetric part. Expanding $(a+b)^2=a^2+b^2+ab+ba$ shows that $a \cdot b$ is a scalar, thereby defining an inner product on $V$. The outer product $a \wedge b$ is a so-called bivector. It can be pictured as a parallelogram with sides $a$ and $b$, and an orientation `from $a$ to $b$'.

To be explicit, let us choose an orthonormal basis $(e_1,\ldots,e_n)$ of the vector space $V$.
The Clifford algebra can be decomposed into a direct sum of linear spaces
\begin{equation}
\mathcal{C\ell}(V^{p,q})
= \oplus_{r=0}^n \mathcal{C\ell}_{r}(V^{p,q}) ,
\end{equation}
where each subspace $\mathcal{C\ell}_r$ is spanned by basis elements of the form $e_{j_1} e_{j_2} \ldots e_{j_r}$. Generic elements of $\mathcal{C\ell}_r$ are referred to as $r$-vectors, or, multivectors of grade $r$. \emph{Simple} $r$-vectors, i.e., $r$-vectors that can be written as an outer product $\blade{a}{r}$ of $r$ individual vectors, are represented geometrically by $r$-dimensional oriented parallelograms with sides $a_1,\ldots,a_r$. Transformations of these geometric objects (such as projections, rotations, or contractions) can be efficiently implemented by operations within the Clifford algebra.\footnote{The basis vectors $e_1,\ldots,e_n$ can be represented by matrices, with the ordinary matrix product taking the role of the Clifford product (cf. the Dirac algebra of $\gamma$-matrices).
However, such representation misses the geometric origin and the geometrical significance of real Clifford algebras. Moreover, any matrix calculation can be performed (more efficiently) using the algebraic rules of Clifford algebras, and a wealth of identities derived therefrom \cite{Hestenes}. }

The subspace of even-grade (or, simply, even) multivectors,
\begin{equation}
\mathcal{C\ell}_{even}(V^{p,q}) 
= \mathcal{C\ell}_0(V^{p,q}) \oplus \mathcal{C\ell}_2(V^{p,q}) \oplus \ldots ,
\end{equation}
is the even subalgebra of $\mathcal{C\ell}(V^{p,q})$. (Indeed, a product of two even multivectors results in an even multivector, owing to property~\eqref{Square} of the Clifford product.) The elements of $\mathcal{C\ell}_{even}$ will be referred to as \emph{real spinors}.

To familiarise with real Clifford algebras, and reveal their geometric significance, we now consider concrete examples of physically relevant vector spaces with increasing dimensionality.

\subsection{Plane}

\begin{figure}
\includegraphics[scale=1]{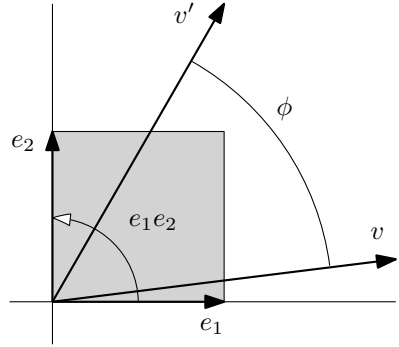}
\caption{Rotation in plane.}
\label{fig:plane}
\end{figure}

Consider a plane $E^2$ spanned by two vectors $e_1$ and $e_2$, with $e_1^2 = e_2^2 = 1$ and $e_1 \cdot e_2 = 0$ (Fig.~\ref{fig:plane}). 
The corresponding real Clifford algebra
\begin{equation}
\mathcal{C\ell}(E^2) 
= span\{1,~e_1,e_2,~e_1e_2\}
\end{equation}
is 4-dimensional. It is a linear span of one scalar $1$, two vectors $e_1$ and $e_2$, and one bivector $e_1 \wedge e_2 = e_1 e_2$.
Calculating $(e_1e_2)^2 = -1$, we realize that the even subalgebra 
\begin{equation}
\mathcal{C\ell}_{even}(E^2)
= span\{1,~e_1e_2\}
\end{equation}
is isomorphic to complex numbers, with the bivector $e_1e_2$ playing the role of the imaginary unit $i$.

The `complex units' $e^{\phi e_1e_2}$, where the exponential is defined via power series in terms of the Clifford product, are of particular geometric significance as they implement rotations. Specifically, a vector $v$ after a rotation by an angle $\phi$ (where the handedness is dictated by the orientation of the bivector $e_1e_2$) is given by\footnote{A proof consists in expanding $v=v_1 e_1 + v_2 e_2$, $e^{\phi e_1e_2} = \cos \phi + e_1e_2 \sin \phi$, and in using the basic identities $e_1e_2=-e_2e_1$ and $e_1^2=e_2^2=1$.}
\begin{equation}
v' = v\, e^{\phi e_1e_2} .
\end{equation}

Let us remark that there is a 1-to-1 correspondence between vectors in $E^2$ and complex numbers, which depends on the choice of the unit vector representing the direction of the `real axis'. For instance, for $e_1$ we have $v \mapsto e_1 v$.

\subsection{Space}

\begin{figure}
\includegraphics[scale=1]{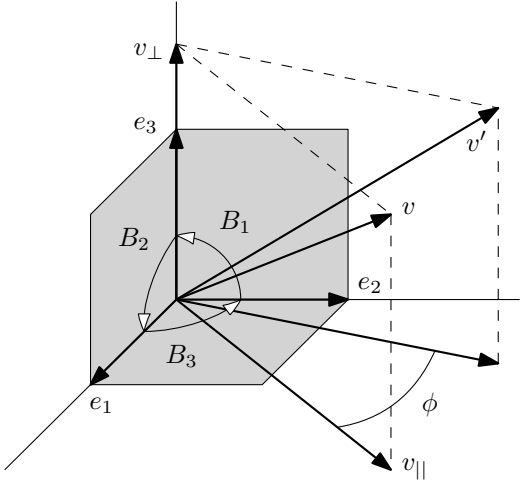}
\caption{Rotation in space with respect to the plane represented by bivector $B_3=e_1 e_2$.}
\label{fig:space}
\end{figure}

Let $e_1,e_2,e_3$, where $e_i\cdot e_j = \delta_{ij}$, be an orthonormal basis of the three-dimensional Euclidean vector space $E^3$ (Fig.~\ref{fig:space}). 
The corresponding real Clifford algebra
\begin{equation}
\mathcal{C\ell}(E^3) 
= span\{ 1,e_1,e_2,e_3,B_1,B_2,B_3,e_1e_2e_3 \}
\end{equation}
is 8-dimensional. It contains three independent bivectors: $B_1=e_2e_3$, $B_2=e_3e_1$, and $B_3=e_1e_2$, each squaring to $-1$. The even subalgebra
\begin{equation}
\mathcal{C\ell}_{even}(E^3)
= span\{1,~B_1,B_2,B_3\}
\end{equation}
is isomorphic to the algebra of quaternions \cite{Hamilton} through the identification $B_1 = -i$, $B_2 = -j$, and $B_3 = -k$.

Rotations in $E^3$ take place with respect to a certain rotation plane (to which there always corresponds, unlike in higher dimensions, a perpendicular rotation axis). Let us consider, for definiteness, the (oriented) plane defined by the bivector $B_3$, i.e., the plane spanned by the basis vectors $e_1$ and $e_2$. A rotation of a vector $v$ by an angle $\phi$ consists in decomposing $v = v_{||} + v_\perp$, and rotating $v_{||}$ (the part parallel to $B_3$) by angle $\phi$, while keeping the perpendicular part $v_\perp$ untouched (see Fig.~\ref{fig:space}). Expressed algebraically,
\begin{equation}
v' = v_{||} e^{\phi B_3} + v_\perp
= e^{-\frac{\phi}{2}B_3}\, v \,e^{\frac{\phi}{2}B_3} ,
\end{equation}
where in the last equality use has been made of the commutation properties $v_{||} B_3 = -B_3 v_{||}$ and $v_\perp B_3 = B_3 v_\perp$.

A generic rotation can be represented as 
\begin{equation} \label{VectorRot}
v'=e^{-B/2}\, v \,e^{B/2} ,
\end{equation}
where the bivector $B$ neatly encodes the rotation plane (together with orientation), and, via its magnitude\footnote{Magnitude of a bivector $B=a \wedge b$ is defined to be the area of the parallelogram spanned by the vectors $a$ and $b$: $|B|=\sqrt{a^2 b^2 - (a \cdot b)^2}$.} $|B|$, the rotation angle. Even-grade Clifford-algebra elements of the form $e^{-B/2} = \cos\frac{|B|}{2} - \frac{B}{|B|}\sin\frac{|B|}{2}$ are called rotors. Under the Clifford product, they form the spin group $Spin(3) \simeq SU(2)$ \cite{Lounesto}.

The two-sided rotor representation of rotations, Eq.~\eqref{VectorRot}, is 2-to-1 degenerate, since $e^{-B/2}$ and $-e^{-B/2}$ yield the same result $v'$. However, it is possible to define a one-sided action
\begin{equation} \label{SpinorRot}
\psi' = e^{-B/2} \psi 
\end{equation}
of rotors on the even subalgebra $\mathcal{C\ell}_{even}(E^3)$ of real spinors $\psi$.

In three dimensions it is always possible to cast a real spinor as $\psi = \sqrt{\rho} \, e^{-B/2}$, for some non-negative scalar $\rho$, and a bivector $B$. This suggests an intuitive picture of real spinors in the three-dimensional space as `rotors with magnitude', in analogy with vectors being thought of as `directions with magnitude'.

\subsection{Spacetime}
\label{sec:Spacetime}

Let $\gamma_0,\gamma_1,\gamma_2,\gamma_3$ be an orthonormal basis of the Minkowski spacetime $E^{1,3}$ ($\gamma_\mu \cdot \gamma_\nu = \eta_{\mu\nu} = diag(1,-1,-1,-1)$). The real Clifford algebra (also referred to as the spacetime algebra \cite{HestenesSTA})
\begin{equation}
\mathcal{C\ell}(E^{1,3}) 
= span\{ 1,~\gamma_\mu,~\gamma_\mu \wedge \gamma_\nu,~\gamma_\mu I,~I \} ,
\end{equation}
where $I = \gamma_0\gamma_1\gamma_2\gamma_3$ denotes the unit pseudoscalar, is 16-dimensional. The even subalgebra
\begin{equation} \label{STEven}
\mathcal{C\ell}_{even}(E^{1,3})
= span\{ 1,~\gamma_\mu \wedge \gamma_\nu,~I \}
\end{equation}
has dimension $8$. It contains 6 independent bivectors that can be arranged into two groups: spatial bivectors $\gamma_2\gamma_1,\gamma_1\gamma_3,\gamma_3\gamma_2$, which square to $-1$; and spatio-temporal bivectors $\gamma_1\gamma_0,\gamma_2\gamma_0,\gamma_3\gamma_0$, which square to $1$.

A (proper orthochronous) Lorentz transformation is described similarly to Eq.~\eqref{VectorRot} in terms of an even multivector (a spacetime rotor) $U$:
\begin{equation} \label{STLorTr}
v' = U v \,\rev{U} 
\quad,\quad
U \rev{U} = 1 .
\end{equation}
Here, the \emph{reversion} $\rev{\,.\,}$ is a linear operation that reverses the order of factors in a Clifford product: $(a \ldots b) \rev{~} = b \ldots a$. In effect, it reverses the sign of bivectors and trivectors while leaving scalars, vectors, and pseudoscalars invariant.

A spacetime rotor $U$ can be cast (up to a possible factor of $-1$) as an exponential of a bivector, which has, in general, three spatial components generating rotations, and three spatio-temporal components generating boosts \cite[Ch.\,5.4]{DoranLas}. Rotors form a Lie group, the spin group $Spin^+(1,3) \simeq SL(2,\mathbb{C})$, and bivectors form the corresponding Lie algebra with the Lie bracket $[B,B'] = B B' - B' B$ \cite[Ch.\,11.3]{DoranLas}.

Real spinors in spacetime are again elements of the even subalgebra.\footnote{It is perhaps worth to note that they can always be decomposed, similarly to the case in three dimensions, as $\Psi = \sqrt{\rho}\, e^{I\beta/2} R$, where $\rho$ is a non-negative scalar, $\beta$ is a scalar, and $R$ is a rotor \cite{Hestenes1967}.\label{PsiForm}} The one-sided action of spacetime rotors on real spinors is completely analogous to Eq.~\eqref{SpinorRot}: $\Psi' = U \Psi$. Apart from this, we can also define the two-sided action $\Psi' = U \Psi \rev{U}$, which will play a prominent role in Sec.~\ref{sec:DiracEq} below.\footnote{The two-sided prescription has, in fact, a very good geometric meaning. Since $$U (v_1 \ldots v_r) \rev{U} = (U v_1 \rev{U}) \ldots (U v_r \rev{U})$$ for any $r$, it acts on a generic multivector by transforming each of its vector constituents according to Eq.~\eqref{STLorTr}.}

\subsection{Generic dimension}

For a generic (pseudo)Euclidean space $E^{p,q}$, $p+q=n$, the real Clifford algebra is a linear space of dimension $2^n$. As before, the real spinors can be defined as the elements of the $2^{n-1}$-dimensional even subalgebra \cite{Francis2005}. Note that this is the smallest linear subspace of $\mathcal{C\ell}(E^{p,q})$, which contains all rotors, and so can naturally serve as a representation space for the one-sided rotor action.

This representation is, in general, not irreducible, for any idempotent element can be used to split the representation space into two invariant subspaces. For example, in the case of the Minkowski spacetime, since $(\gamma_3\gamma_0)^2=1$, we have two even multivectors $(1 \pm \gamma_3\gamma_0)/2$, which act by right multiplication as the corresponding projectors \cite[Ch.\,17]{Lounesto}.

\section{Relating real and complex spinors}
\label{sec:Relating}

The spinors used in quantum theory are most commonly defined as complex vectors, acted upon by complex matrices. In this section we demonstrate the capability of real spinors to substitute the traditional complex spinors by establishing a 1-to-1 correspondence between the two systems. This is possible in space and in spacetime, since the dimensions of their respective even subalgebras match the number of \emph{real} components of the $\mathbb{C}^2$ and $\mathbb{C}^4$ vectors.

\subsection{Pauli spinors and real spinors in space}
\label{sec:RelatingPauli}

In the three-dimensional space $E^3$, the 4-(real)-component real spinors $\psi$ are in 1-to-1 correspondence with Pauli spinors
\begin{equation}
\ket{\psi} = 
\begin{pmatrix}
z_0 \\ z_1
\end{pmatrix}
\in \mathbb{C}^2 ,
\end{equation}
where the two complex components of $\ket{\psi}$ are defined (following Ref.~\cite{Doran1993})
\begin{align} \label{PauliDef}
z_0 &= \avg{\psi(1-i \,B_3)} , 
\nonumber\\
z_1 &= \avg{B_2 \psi(1-i \,B_3)} .
\end{align}
Here, the operation $\avg{\,.\,}$ singles out the scalar part of a multivector. In a matrix representation of the Clifford algebra this corresponds to the (normalized) trace of a matrix. For any two multivectors, it enjoys the commutation property $\avg{A B} = \avg{B A}$.

It is straightforward to translate between Clifford-algebra operations on real spinors $\psi$, and matrix operations on their $\mathbb{C}^2$ counterparts $\ket{\psi}$. The structure of definition \eqref{PauliDef}, which may have seemed quite arbitrary, in fact ensures that
\begin{equation} \label{3Dleft}
\ket{B_j \psi}
= i \hat{\sigma}_j \ket{\psi} ,
\end{equation}
where $\hat{\sigma}_j$ are the Pauli matrices
\begin{equation}
\hat{\sigma}_1 = 
\begin{pmatrix}
0 & 1 \\ 1 & 0
\end{pmatrix}
,~
\hat{\sigma}_2 = 
\begin{pmatrix}
0 & -i \\ i & 0
\end{pmatrix}
,~
\hat{\sigma}_3 = 
\begin{pmatrix}
1 & 0 \\ 0 & -1
\end{pmatrix} .
\end{equation}

Expressions for right-side bivector multiplications,
\begin{subequations} \label{3Dright}
\begin{align}
\ket{\psi B_1}
&= \hat{\sigma}_2 \ket{\psi}^* ,
\label{3Dright1}\\
\ket{\psi B_2}
&= i\hat{\sigma}_2 \ket{\psi}^* ,
\label{3Dright2}\\
\ket{\psi B_3}
&= i \ket{\psi} ,
\label{3Dright3}
\end{align}
\end{subequations}
involve complex conjugation $\ket{\psi}^* = (z_0^*,z_1^*)^T$. 
The last of these relations provides a geometric interpretation of the formal imaginary unit $i$ of the Pauli matrix theory. It asserts that $i$ is equivalent to the bivector $B_3$ acting on the right of the real spinors. 

Finally, let us remark that the standard scalar product on $\mathbb{C}^2$, $\braket{\psi}{\psi'} = z_0^* z_0' + z_1^* z_1'$, is represented on real spinors by
\begin{equation}
\braket{\psi}{\psi'}
= \avg{\rev{\psi} \psi' (1 - i\,B_3)} .
\end{equation}
In particular, $\braket{\psi}{\psi} = \avg{\rev{\psi} \psi}$, since the bivector $\psi B_3 \rev{\psi}$ is erased by the scalar part operation.

\subsection{Dirac spinors and real spinors in spacetime}

In Minkowski spacetime $E^{1,3}$, the 8-(real)-component real spinors $\Psi$ are in 1-to-1 correspondence with Dirac spinors
\begin{equation}
\ket{\Psi} = 
\begin{pmatrix}
z_0 \\ z_1 \\ z_2 \\ z_3
\end{pmatrix}
\in \mathbb{C}^4 ,
\end{equation}
where the four complex components of $\ket{\Psi}$ are defined
\begin{align} \label{DiracDef}
z_0 &= \avg{\Psi(1-i \gamma_2\gamma_1)} , 
\nonumber\\
z_1 &= \avg{\gamma_1\gamma_3 \Psi(1-i \gamma_2\gamma_1)} ,
\nonumber\\
z_2 &= \avg{\gamma_3\gamma_0 \Psi(1-i \gamma_2\gamma_1)} ,
\nonumber\\
z_3 &= \avg{\gamma_1\gamma_0 \Psi(1-i \gamma_2\gamma_1)} .
\end{align}

The Dirac matrices now enter the scene as matrix equivalents of the Clifford algebra operations $\Psi \mapsto \gamma_\mu \Psi \gamma_0$, where the right multiplication by $\gamma_0$ ensures that the result stays in the even subalgebra. It is straightforward to verify that 
\begin{equation} \label{GammaMat}
\ket{\gamma_\mu \Psi \gamma_0}
= \hat{\gamma}_\mu \ket{\Psi} ,
\end{equation}
where
\begin{align}
\hat{\gamma}_0 = 
\left(\begin{smallmatrix}
1 & ~ & ~ & ~ \\
~ & ~1 & ~ & ~ \\
~ & ~ & -1 & ~ \\
~ & ~ & ~ & -1
\end{smallmatrix}\right)
,~
\hat{\gamma}_1 =
\left(\begin{smallmatrix}
~ & ~ & ~ & -1 \\ 
~ & ~ & -1 & ~ \\ 
~ & ~1 & ~ & ~ \\ 
1 & ~ & ~ & ~
\end{smallmatrix}\right)
, \nonumber\\
\hat{\gamma}_2 =
\left(\begin{smallmatrix}
~ & ~ & ~ & ~i \\ 
~ & ~ & -i & ~ \\ 
~ & -i & ~ & ~ \\ 
i & ~ & ~ & ~
\end{smallmatrix}\right)
,~
\hat{\gamma}_3 =
\left(\begin{smallmatrix}
~ & ~ & -1 & ~ \\ 
~ & ~ & ~ & ~1 \\ 
1 & ~ & ~ & ~ \\ 
~ & -1 & ~ & ~
\end{smallmatrix}\right) 
\end{align}
denote the Dirac $\gamma$-matrices in the standard Dirac representation \cite{Itzykson}.\footnote{Of course, the reason we recover Dirac matrices in this particular representation is due to a particularity of definitions~\eqref{DiracDef}. The various representations of $\gamma$-matrices can be obtained by appropriate redefinitions of the complex components $z_0,z_1,z_2,z_3$ \cite{Doran1993}.} Since $\gamma_0^2 = 1$, it follows readily that 
\begin{equation} \label{LeftBivector}
\ket{\gamma_\mu \gamma_\nu \Psi} 
= \hat{\gamma}_\mu \hat{\gamma}_\nu \ket{\Psi} .
\end{equation}

For later reference, let us quote the results of right-side multiplications by the 6 bivectors of the spacetime Clifford algebra:
\begin{subequations} \label{RightBivector}
\begin{align} 
\ket{\Psi \gamma_1 \gamma_0}
&= -i \hat{\gamma}_2 \ket{\Psi}^* ,
\label{RightBivector10}\\
\ket{\Psi \gamma_2 \gamma_0}
&= \hat{\gamma}_2 \ket{\Psi}^* ,
\label{RightBivector20}\\
\ket{\Psi \gamma_3\gamma_0}
&= \hat{\gamma}_5 \ket{\Psi} ,
\label{RightBivector30}\\
\ket{\Psi \gamma_3 \gamma_2}
&= -\hat{\gamma}_2 \hat{\gamma}_5 \ket{\Psi}^* ,
\label{RightBivector32}\\
\ket{\Psi \gamma_1 \gamma_3}
&= -i \hat{\gamma}_2 \hat{\gamma}_5 \ket{\Psi}^*  ,
\label{RightBivector13}\\
\ket{\Psi \gamma_2 \gamma_1}
&= i \ket{\Psi} ,
\label{RightBivector21}
\end{align}
\end{subequations}
where $\ket{\Psi}^* = (z_0^*,z_1^*,z_2^*,z_3^*)^T$, and we denoted $\hat{\gamma}_5 = i\hat{\gamma}^0\hat{\gamma}^1\hat{\gamma}^2\hat{\gamma}^3$ as common. The last of these equations shows that multiplication of Dirac spinors by $i$ translates into right multiplication of real spinors by the spatial bivector $\gamma_2\gamma_1$.

The standard scalar product on the space of Dirac spinors features the Dirac conjugation $\brast{\bar{\Psi}} = \bra{\Psi} \hat{\gamma}_0$, and is represented on the space of real spinors by 
\begin{equation}
\braket{\bar{\Psi}}{\Psi'} = \avg{\rev{\Psi}\Psi'(1-i\gamma_2\gamma_1)} .
\end{equation}
The scalar product on spinors allows one to construct bilinear covariants of the Dirac theory, and express them in terms of real spinors \cite[Ch.\,8.2]{DoranLas}. The most prominent of these observables is the (vector-valued) Dirac current
\begin{equation}
\Psi \gamma_0 \rev{\Psi}
= \brast{\bar{\Psi}} \hat{\gamma}^\mu \ket{\Psi} \gamma_\mu ,
\end{equation}
(Indices are raised and lowered using the Minkowski metric.)

Finally, let us remark that the algebra of real spinors is endowed with an invertible product $\Psi \Psi'$ (to justify, see Footnote~\ref{PsiForm}). This structure is missing in the complex vector space $\mathbb{C}^4$, but can be induced by defining $\circ: \mathbb{C} \times \mathbb{C} \rightarrow \mathbb{C}$, $\ket{\Psi} \circ \ket{\Psi'} = \ket{\Psi\Psi'}$. (Analogous remark holds in the case of spatial real spinors and the complex space $\mathbb{C}^2$.)

\section{Real Dirac equation}
\label{sec:DiracEq}

With the mathematical formalism of previous sections in hand, we are in position to apply the real Clifford algebras in the quantum theory. Namely, in this article, we investigate the Dirac equation of relativistic quantum mechanics, which is more fundamental than the non-relativistic Pauli or Schr\"{o}dinger equations, and, at the same time, it is easier to study, as it contains only first-order derivatives.

The traditional `matrix' Dirac equation \cite{Dirac1928} (with $\hbar = c = 1$)
\begin{equation} \label{DiracEq}
i\hat{\gamma}^\mu \partial_\mu \ket{\Psi} - m \ket{\Psi} = 0
\end{equation}
can be cast using Eqs.~\eqref{GammaMat} and \eqref{RightBivector}
in the real-Clifford-algebra form
\begin{equation} \label{HestenesEq}
\gamma^{\mu} \partial_\mu \Psi \gamma_0 \gamma_2 \gamma_1 - m \Psi = 0 , 
\end{equation}
found by Hestenes in \cite{Hestenes1967}. The latter equation, including the electromagnetic coupling term, has been further analysed, e.g., in Refs.~\cite{DoranLas,Hestenes1975,Hestenes2003,Hestenes1982,Hestenes2008,Doran2005}.

The conceptual value of Eq.~\eqref{HestenesEq} lies in the fact that it completely eliminates the uninterpreted formal imaginary unit $i$, present in the original Dirac's equation~\eqref{DiracEq}. On the other hand, a disturbing feature of the real formulation is the presence of the term $\gamma_0 \gamma_2 \gamma_1$, which seems to break Lorentz invariance. It is therefore important to clarify in which sense we regard Eq.~\eqref{HestenesEq} as invariant under Lorentz transformations.

\subsection{Lorentz transformations on flat spacetime}

A Lorentz transformation $x'^\mu = \Lambda^\mu_{~\,\nu} x^\nu$ alters the left-hand side of Eq.~\eqref{HestenesEq} to
\begin{equation} \label{DiracTransf}
\gamma'^\mu (\Lambda^{-1})^\nu_{~\,\mu} \partial_\nu \Psi' \gamma'_0 \gamma'_2 \gamma'_1 - m \Psi' .
\end{equation}
For this to yield zero, we need to prescribe appropriate transformation properties for the quantities $\gamma_\mu$ and $\Psi$.

First, let us observe that the vectors $(\Lambda^{-1})^\nu_{~\,\mu} \gamma^\mu$ form an orthonormal frame, which can be expressed in terms of the original frame $\gamma^\nu$:
\begin{equation}
(\Lambda^{-1})^\nu_{~\,\mu} \gamma^\mu
= U \gamma^\nu \rev{U} ,
\end{equation}
for some spacetime rotor $U$ (see Eq.~\eqref{STLorTr}).

The standard method by which Lorentz invariance of the Dirac equation is established (e.g., in Ref.~\cite[Ch.\,3]{GreinerRQM}) proceeds by postulating the transformation rules
\begin{equation} \label{OneSidedTr}
\gamma'_\mu = \gamma_\mu 
\quad,\quad
\Psi'(x') = U \Psi (x) .
\end{equation}
That is, the spinor $\Psi$ is transformed by the one-sided action of $U$, while the frame vectors $\gamma_\mu$ undergo no change at all.\footnote{In Ref.~\cite[p.\,45]{FeynmanQED}, another possibility is coined, which consists in transforming $\gamma'_\mu = \rev{U} \gamma_\mu U$, and postulating a scalar transformation law for the spinor, $\Psi'(x')=\Psi(x)$. The term $\gamma'_0 \gamma'_2 \gamma'_1 = \rev{U} \gamma_0 \gamma_2 \gamma_1 U$ in Eq.~\eqref{DiracTransf} would then, however, impede the desired Lorentz invariance.} This is at odds with our definition of real spinors as even multivectors in the real Clifford algebra, for multivectors are polynomials in the basis vectors $\gamma_\mu$, and as such should have complementary transformation properties.

To resolve this tension we consider a general curved-spacetime setting (of which the flat spacetime is a special case).

\subsection{Curved spacetime and local Lorentz transformations}

\begin{figure}
\includegraphics[scale=1]{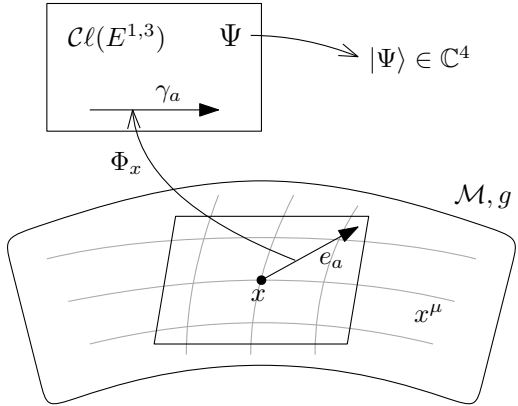}
\caption{Isometries $\Phi_x$ map tangent spaces of the spacetime manifold onto a `model' Minkowski space $E^{1,3}$. The even part of the real Clifford algebra $\mathcal{C}\ell(E^{1,3})$ accommodates real spinors $\Psi$, which can be translated to $\mathbb{C}^4$ by the mapping $\ket{\,.\,}$ defined in Eq.~\eqref{DiracDef}. }
\label{fig:curved}
\end{figure}

Let the spacetime be a Lorentzian manifold $\mathcal{M}$ with metric tensor $g_{\mu\nu}$, and choose a tetrad of orthonormal vectors fields $e_a = e_a^\mu \partial_\mu$, $a=0,1,2,3$, satisfying $g_{\mu\nu} e_a^\mu e_b^\nu = \eta_{a b}$. Each tangent space of the manifold is (non-canonically) isomorphic to a `model' Minkowski vector space $E^{1,3}$ via a linear isometry $\Phi_x(e_a) = \gamma_a$ ($\gamma_a \cdot \gamma_b = \eta_{a b}$). See Fig.~\ref{fig:curved}.

There are two kinds of transformations to be consider on a curved spacetime \cite[Ch.\,13]{WaldGR}: 1) coordinate changes $x^\mu \rightarrow x'^\mu$; and 2) local Lorentz (gauge) transformations, which, in our setting, alter the isomorphisms (or, the gauge) $\Phi_x$ so that
\begin{equation} \label{RealGammaTr}
\gamma'_a = \Phi'_x(e_a) = U \gamma_a \rev{U}
\end{equation}
for some rotor field $U(x)$.
The latter induce corresponding transformations on the entire Clifford algebra $\mathcal{C}\ell(E^{1,3})$, in particular, the real spinor fields $\Psi(x)$ transform as
\begin{equation} \label{RealSpTr}
\Psi' = U \Psi \rev{U} .
\end{equation}

A modified version of Eq.~\eqref{HestenesEq} invariant under both kinds of transformations 1) and 2) reads
\begin{equation} \label{HestenesEqCurved}
\gamma^{a} e^\mu_a (D_\mu \Psi) \gamma_0 \gamma_2 \gamma_1 - m \Psi = 0 , 
\end{equation}
where $\gamma^a = \eta^{a b} \gamma_b$, and the covariant derivative
\begin{equation}
D_\mu \Psi
= \partial_\mu \Psi - \omega_\mu \Psi + \Psi \mathcal{A}_\mu
\end{equation}
features a pair of $4 \times 6$-component gauge fields $\omega_\mu$ and $\mathcal{A}_\mu$ (bivector-valued 1-forms) with identical transformation properties
\begin{align} \label{GaugeFieldTr}
\omega'_\mu  
&= U \omega_\mu \rev{U} + (\partial_\mu U) \rev{U} ,
\nonumber\\
\mathcal{A}'_\mu 
&= U \mathcal{A}_\mu \rev{U} + (\partial_\mu U) \rev{U} .
\end{align}
In the sequel we will see that $\omega_\mu$ represents the spin connection (so far with no a priori relation to the tetrad $e_a^\mu$), while $\mathcal{A}_\mu$, in essence, generalizes the electromagnetic gauge potential $A_\mu$.

First, observe that local Lorentz transformations can be used to set the basis vector fields $\gamma_a$ to constants. In this gauge it is straightforward to use formulas \eqref{GammaMat}, \eqref{LeftBivector} and \eqref{RightBivector21} to convert Eq.~\eqref{HestenesEqCurved} to the matrix form
\begin{equation}
i \hat{\gamma}^{a} e^\mu_a \ket{D_\mu \Psi} - m \ket{\Psi} = 0 , 
\end{equation}
where the covariant derivative translates as
\begin{equation}
\ket{D_\mu \Psi} 
= \partial_\mu \ket{\Psi} + \frac{\omega_\mu^{a b}}{2} \hat{\gamma}_a \hat{\gamma}_b \ket{\Psi} - \frac{\mathcal{A}_\mu^{a b}}{2} \ket{\Psi \gamma_a \gamma_b} .
\end{equation}
Here, $\omega_\mu^{a b} = \omega_\mu \cdot (\gamma^a \wedge \gamma^b)$ are the components of the field $\omega_\mu = \frac{1}{2}\omega_\mu^{a b} \gamma_b \gamma_a$, and accordingly for $\mathcal{A}_\mu$. Furthermore, we can utilise Eq.~\eqref{RightBivector21} to identify the electromagnetic potential $A_\mu$ as the part of $\mathcal{A}_\mu$ corresponding to the bivector $\gamma_2\gamma_1$, namely, 
\begin{equation}
e A_\mu
= \mathcal{A}_\mu^{1 2} = -\mathcal{A}_\mu^{2 1} .
\end{equation}

Interpretation of the other parts of $\mathcal{A}_\mu$ is, however, not very clear. We may only note at this point that the $\gamma_3 \gamma_0$-part, Eq.~\eqref{RightBivector30}, is related to chirality, and the $\gamma_2 \gamma_0$-part, Eq.~\eqref{RightBivector20}, to charge conjugation \cite{Itzykson}.

Several further comments are in order. According to Eq.~\eqref{GaugeFieldTr}, $\omega_\mu$ and $\mathcal{A}_\mu$ have the same transformation properties, and they differ only by the side of $\Psi$ on which they act. We could, in fact, identify $\omega_\mu = \mathcal{A}_\mu$ while still preserving local Lorentz invariance of Eq.~\eqref{HestenesEqCurved}. However, the identification of $\omega_\mu$ with a spin connection, and $\mathcal{A}_\mu$ with a generalized (non-Abelian) electromagnetic potential suggests that they be treated as a priori distinct quantities.

In spite of the unusual `two-sided' form of the transformation rule~\eqref{RealSpTr} for real spinors, observable quantities are invariant under local Lorentz transformations. Indeed, for example, the components of the Dirac current
\begin{equation}
J^\mu 
= e_a^\mu \brast{\bar{\Psi}} \hat{\gamma}^a \ket{\Psi} 
= e_a^\mu \avg{\gamma^a \Psi \gamma_0 \rev{\Psi}}
\end{equation} 
are scalars in the Clifford algebra $\mathcal{C}\ell(E^{1,3})$.

In literature on the real Dirac equation (e.g., in Ref.~\cite[Ch.\,13]{DoranLas}) it is common to adopt (a local version of) the traditional transformation rules~\eqref{OneSidedTr} rather than Eqs.~\eqref{RealGammaTr} and \eqref{RealSpTr}. Then, transformations of the form $\Psi \rightarrow \Psi e^M$ that preserve the real Dirac Lagrangian (or, alternatively, the Dirac current), namely, such that $M \in span\{ \gamma_2\gamma_1,\gamma_1\gamma_3,\gamma_3\gamma_2, I \}$, are found and gauged. This leads to certain insights regarding the structure of the electroweak theory \cite{Hestenes1982,Hestenes2008}.

\subsection{Gauge field strengths and further observations}

To determine the field strengths corresponding to the gauge fields $\omega_\mu$ and $\mathcal{A}_\mu$, we form the commutator of covariant derivatives. A straightforward calculation yields
\begin{equation}
[D_\mu , D_\nu] \Psi
= -\mathcal{R}_{\mu\nu} \Psi + \Psi \mathcal{F}_{\mu\nu} ,
\end{equation}
where 
\begin{align}
\mathcal{R}_{\mu\nu}
&= \partial_\mu \omega_\nu - \partial_\nu \omega_\mu - [\omega_\mu,\omega_\nu] ,
\nonumber\\
\mathcal{F}_{\mu\nu}
&= \partial_\mu \mathcal{A}_\nu - \partial_\nu \mathcal{A}_\mu - [\mathcal{A}_\mu,\mathcal{A}_\nu] 
\end{align}
are bivector-valued 2-forms. For the purely electromagnetic case $\mathcal{A}_\mu = e A_\mu \gamma_2 \gamma_1$, we find $\mathcal{F}_{\mu\nu} = e F_{\mu\nu} \gamma_2 \gamma_1$ with $F_{\mu\nu} = \partial_\mu A_\nu - \partial_\nu A_\mu$, as expected.

Finally, let us point out several observations that could be relevant for further studies.
(From now on we restrict ourselves, for simplicity, to flat spacetime: $\gamma^a e^\mu_a = \gamma^\mu = const.$ and $\omega_\mu = 0$.)

{\it Remark 1}: 
In analogy with electromagnetism one could propose a kinetic term for the field $\mathcal{A}_\mu$ of the form $\avg{\mathcal{F}_{\mu\nu} \mathcal{F}^{\mu\nu}}$, and derive the equations of motion
\begin{equation} \label{YMeq}
\partial_\mu \mathcal{F}^{\mu\nu} 
- [\mathcal{A}_\mu, \mathcal{F}^{\mu\nu}] 
= 0 .
\end{equation}
These are non-linear partial differential equations of a Yang-Mills type, which reduce to Maxwell equations when $\mathcal{A}_\mu = e A_\mu \gamma_2 \gamma_1$. Note that although the introduction of the non-Abelian gauge field $\mathcal{A}_\mu$ was based on our study of real spinors and real Dirac equation, Eq.~\eqref{YMeq}, as it stands, is independent of the notion of spinors.

{\it Remark 2}: The real Dirac equation~\eqref{HestenesEqCurved} multiplied on the right by $\gamma_2 \gamma_1 \rev{\Psi}$ takes the form
\begin{equation}
\gamma^{\mu} (\partial_\mu \Psi) \gamma_0 \rev{\Psi} 
- \gamma^\mu \Psi \frac{\mathcal{A}_\mu}{2} \gamma_0 \rev{\Psi}
+ m \Psi \gamma_2 \gamma_1 \rev{\Psi} = 0 .
\end{equation}
Adding the reverse of this equation, and taking the scalar part yields
\begin{equation}
\gamma^\mu \cdot \big( \partial_\mu (\Psi \gamma_0 \rev{\Psi}) \big) 
= \gamma^\mu \cdot \big( \Psi (\mathcal{A}_\mu \cdot \gamma_0) \rev{\Psi} \big) .
\end{equation}
That is, the current $J^\mu = \gamma^\mu \cdot (\Psi \gamma_0 \rev{\Psi})$ is not conserved unless $\mathcal{A}_\mu$ are purely spatial bivectors (as is the case of electromagnetism).

{\it Remark 3}: Footnote~\ref{PsiForm} implies that non-zero real spinors have inverse $\Psi^{-1}$. This fact can be used to cast Eq.~\eqref{HestenesEqCurved} (for the purely electromagnetic case $\mathcal{A}_{\mu} = e A_\mu \gamma_2 \gamma_1$) as
\begin{equation}
\gamma^\mu (\partial_\mu \Psi) \gamma_2 \gamma_1 \Psi^{-1}
+ q \gamma^\mu A_\mu 
- m \Psi \gamma_0 \Psi^{-1} = 0 .
\end{equation}
From this the potential $A_\mu$ can be readily expressed in terms of the field $\Psi$ and its derivatives (this observation was made earlier in \cite{Radford1996} using the 2-spinor formalism).

\section{Conclusion}

In this article we reviewed the construction of spinors as even multivectors in Clifford algebras over real vector spaces. Unlike in the traditional approach, where spinors are regarded as elements of complex vector spaces, acted upon by complex matrices, the real spinors $\Psi$ inherit the Clifford product structure, and the operations, expressed by multiplication by Clifford algebra elements, can take place on both the left and the right side of $\Psi$.
 
In fact, in view of definitions~\eqref{DiracDef}, it is tempting to regard the real spinors as more `real' (fundamental) than the complex spinors, as the former are more closely related to the geometry of spacetime, and the representation by 4 complex numbers is merely a way to redistribute the 8 real degrees of freedom of the real spinors in a geometrically somewhat less transparent manner.

Although we focused on real spinors in spacetime and on the Dirac equation of relativistic quantum mechanics, real spinors find applications also in non-relativistic quantum mechanics  \cite{Hestenes1971}, \cite[Ch.\,8]{DoranLas}. Indeed, the real versions of the Sch\"{o}dinger and Pauli equations can be obtained rather easily using the relations of Sec.~\ref{sec:RelatingPauli}.

Comparing the real and the complex formulations of the Dirac equation on aesthetic grounds, it may seem that the complex matrix formulation, Eq.~\eqref{DiracEq}, is more elegant (symmetric) than the real one, Eq.~\eqref{HestenesEq}, which involves the explicit factor $\gamma_0 \gamma_2 \gamma_1$. Rather than a reason to discard the real formulation, we regard this aesthetic point as an encouragement for seeking an equation that would be more refined (and perhaps more fundamental).

The main contribution of the present study consists in the introduction of the non-Abelian gauge field $\mathcal{A}_\mu$, which takes values in the algebra of spacetime bivectors ($\simeq \mathfrak{so}(1,3)$). It was deduced from the requirement of local Lorentz invariance, and the assumption that Lorentz transformations act in the same way on all elements of the Clifford algebra, including the vectors $\gamma_a$ and the real spinors $\Psi$. To decide whether the non-electromagnetic parts of $\mathcal{A}_\mu$ are physically significant quantities, one would have to carry out further analysis going beyond the scope of this article, and perhaps beyond the realm of relativistic quantum mechanics towards the quantum field theory. The present article can be thought of as a starting point for such endeavour.


\section*{Acknowledgements}

The author would like to acknowledge stimulating discussions with Leslaw Rachwal, Josef Schmidt, Jan Vysok\'{y}, and Alejandro Perez.

\appendix

%



\end{document}